\documentclass[aps, prd, nofootinbib, preprintnumbers, nobibnotes, reprint, twocolumn]{revtex4-1}

\usepackage{amssymb}
\usepackage{amsmath}
\usepackage{graphicx,subfigure}
\usepackage{longtable}
\usepackage{verbatim}
\usepackage{amsfonts}
\usepackage{hyperref}
\usepackage{cancel}
\usepackage{braket}
\usepackage{bbm}
\usepackage{slashed}
\usepackage{gensymb}
\usepackage{color}

\newcommand{\ie}{\emph{i.e.}}

\newcommand{\be}{\begin{equation}}
\newcommand{\ee}{\end{equation}}
\newcommand{\bea}{\begin{eqnarray}}
\newcommand{\eea}{\end{eqnarray}}
\newcommand{\beq}{\begin{equation}}
\newcommand{\eeq}{\end{equation}}
\def\beqa{\begin{eqnarray}}
  \def\eeqa{\end{eqnarray}}
\newcommand{\bv}{\left(\begin{array}{c}}
\newcommand{\ev}{\end{array}\right)}

\def\lsim{\mathrel{\rlap{\lower4pt\hbox{\hskip1pt$\sim$}}
    \raise1pt\hbox{$<$}}}         
\def\gsim{\mathrel{\rlap{\lower4pt\hbox{\hskip1pt$\sim$}}
    \raise1pt\hbox{$>$}}}         
    
\newcommand{\nn}{\nonumber}

\newcommand{\imag}{\mathrm{Im}\,}
\newcommand{\eq}[1]{Eq.~(\ref{#1})}

\newcommand{\fig}[1]{Fig.~\ref{#1}}
\newcommand{\ov}{\overline}

\newcommand{\mev}{\,\mbox{MeV}}
\newcommand{\Dbar}{\,\overline{\!D}}
\newcommand{\DorDbar}{\raisebox{7.7pt}{$\scriptscriptstyle(\hspace*{8.5pt})$}
  \hspace*{-10.7pt}\!\Dbar}

\begin{document}

\preprint{TTP17-033}

\title{Neutral $D\rightarrow K K^*$ decays as discovery channels for charm CP violation} 

\author{Ulrich Nierste$^{\,a}$}
\email{ulrich.nierste@kit.edu}
\author{Stefan Schacht$^{\,b}$}
\email{schacht@to.infn.it }

\affiliation{
$^{\,a}$ Institut f\"ur Theoretische Teilchenphysik, Karlsruher
  Institut f\"ur Technologie, 76128 Karlsruhe, Germany\\
$^{\,b}$ Dipartimento di Fisica, Universit\`a di Torino \& INFN, Sezione di Torino, I-10125 Torino, Italy
}

\vspace*{1cm}

\begin{abstract}
We point out that the CP asymmetries in the decays $D^0\rightarrow K_{S} K^{*0}$ and $D^0\rightarrow K_{S} \overline{K}{}^{*0}$
are potential discovery channels for charm CP violation in the Standard Model. 
We stress that no flavor tagging is necessary, the untagged CP asymmetry 
$a_{CP}^{\mathrm{dir}}( \raisebox{7.7pt}{$\scriptscriptstyle(\hspace*{8.5pt})$} \hspace*{-10.7pt}\!\,\overline{\!D} \to K_S K^{*0})$
is essentially equal to the tagged one, so that the untagged measurement comes with a significant statistical gain.
Depending on the relevant strong phase, $\vert a_{CP}^{\text{dir, untag}}\vert$ can be as large as $0.003$.
The CP asymmetry is dominantly generated by exchange diagrams and does not require non-vanishing penguin amplitudes.
While the CP asymmetry is smaller than in the case of $D^0\rightarrow K_SK_S$, the experimental analysis is more
efficient due to the prompt decay $K^{*0}\rightarrow K^+\pi^-$. One may further search for favourable strong phases
in the Dalitz plot in the vicinity of the $K^{*0}$ peak.
\end{abstract}

\maketitle


\section{Introduction \label{sec:intro}}
Charm CP violation has not been discovered yet. Within the Standard
Model (SM) all CP asymmetries involve the combination $\lambda_b\equiv
V_{cb}^*V_{ub}$ of elements of the Cabibbo-Kobayashi-Maskawa (CKM)
matrix. The smallness of $|\lambda_b|$ had nurtured the hope that  
new physics would manifest itself in orders-of-magnitude enhancements 
of CP asymmetries.
However, this scenario is seemingly not realized in nature, so that the
scientific goals to discover charm CP violation and to establish new
physics involve distinct strategies.  In this paper we address the first
topic and discuss how charm CP violation can be discovered best,
assuming that there is only the SM contributions governed by
$\lambda_b$.

Singly Cabibbo-suppressed (SCS) decay amplitudes of $D$ mesons 
involve the CKM elements $\lambda_q\equiv V_{cq}^*V_{uq}$ with $q=d,s$
or $b$. Using $\lambda_d+\lambda_s+\lambda_b=0$ one may express the 
amplitude of some decay $d$ as
\begin{align}
\mathcal{A}(d) &\equiv \lambda_{sd} \mathcal{A}_{sd}(d) - 
    \frac{\lambda_b}{2} \mathcal{A}_b(d)\,,  \label{eq:amp-ckm-decomposition}  
\end{align} 
with $\lambda_{sd} = (\lambda_s-\lambda_d)/2$. Branching ratios are
completely dominated by the first term $\lambda_{sd}
\mathcal{A}_{sd}(d)$.  The direct CP asymmetry reads 
\begin{align}
a_{CP}^{\mathrm{dir}}(d) &\equiv 
\frac{\vert \mathcal{A}(d)\vert^2 - \vert \overline{\mathcal{A}}(d)\vert^2}{\vert \mathcal{A}(d)\vert^2 + \vert \overline{\mathcal{A}}(d)\vert^2} \label{eq:aCP-def} \\
&=  \imag \frac{\lambda_b}{\lambda_{sd}} \imag 
   \frac{\mathcal{A}_b(d)}{\mathcal{A}_{sd}(d)} . \label{eq:aCPdirFormula}  
\end{align}
$\mathcal{A}_{sd}(d)$ and $\mathcal{A}_b(d)$ can be written as the sum
of different topological amplitudes; in the limit of exact flavor-SU(3)
symmetry these are the \emph{tree (T)}, \emph{color-suppressed tree
  (C)}, \emph{exchange (E)}, \emph{annihilation (A)}, \emph{penguin
  (P$_q$)}, and \emph{penguin annihilation (PA$_q$)} amplitudes. The
latter two topologies involve a loop with the indicated internal quark
$q=d,s,b$. In essentially all commonly studied decays (including the
popular modes $D^0\to \pi^+\pi^-$ and $D^0\to K^+ K^-$)
$\mathcal{A}_b(d)/\mathcal{A}_{sd}(d)$ is proportional to $P\equiv P_s
+P_d -2 P_b$. 

\begin{figure*}[t]
\begin{center}
\subfigure[\label{}]{
        \includegraphics[width=0.23\textwidth]{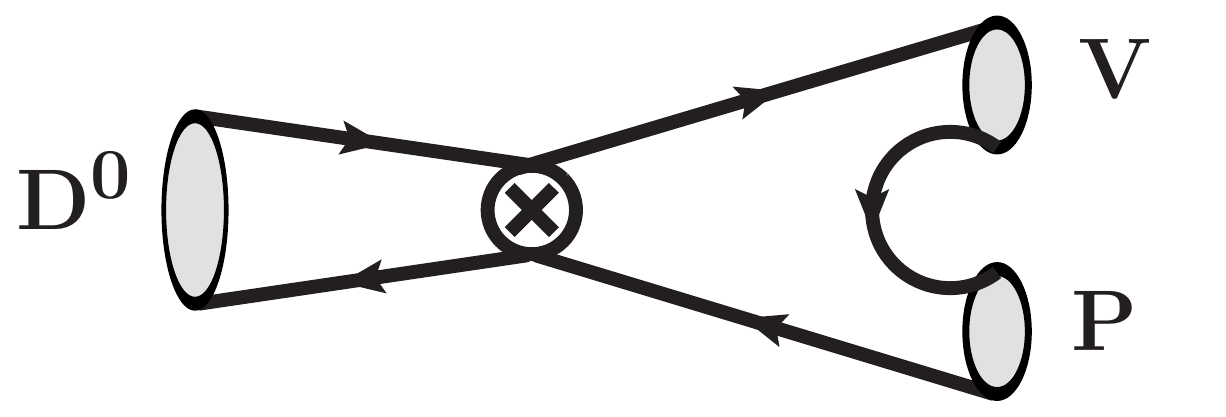}
}
\hfill %
\subfigure[\label{}]{
        \includegraphics[width=0.23\textwidth]{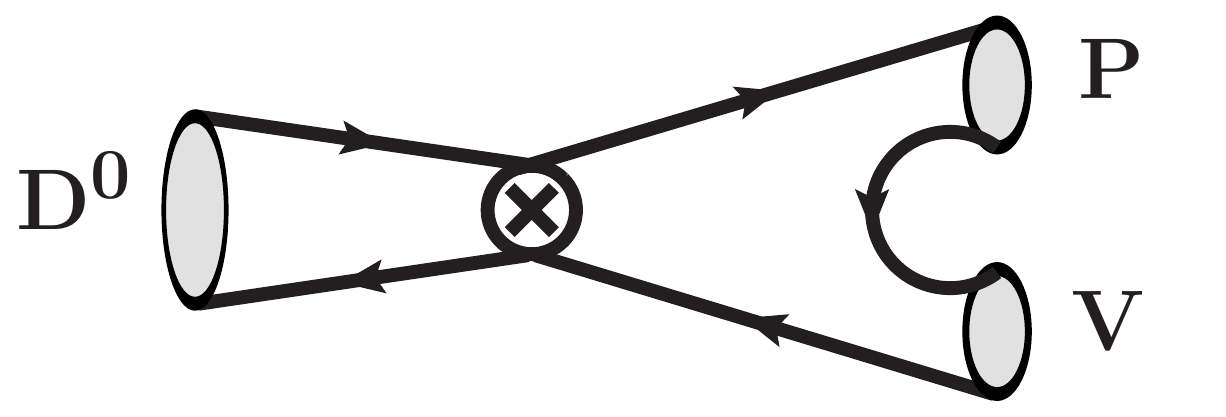}
}
\hfill %
\subfigure[\label{}]{
        \includegraphics[width=0.23\textwidth]{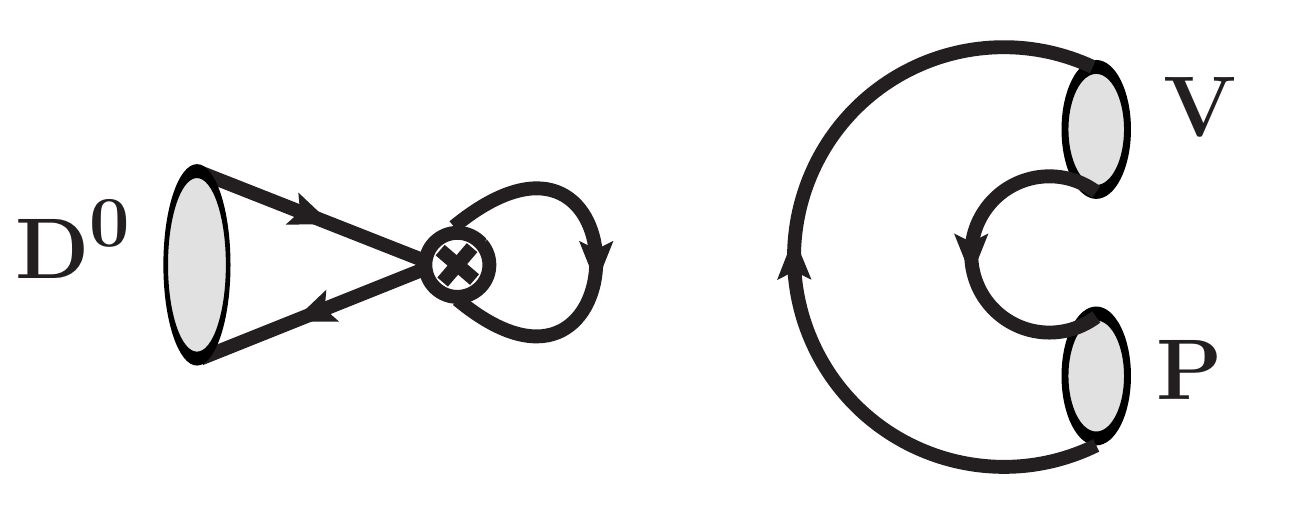}
}
\hfill %
\subfigure[\label{}]{
        \includegraphics[width=0.23\textwidth]{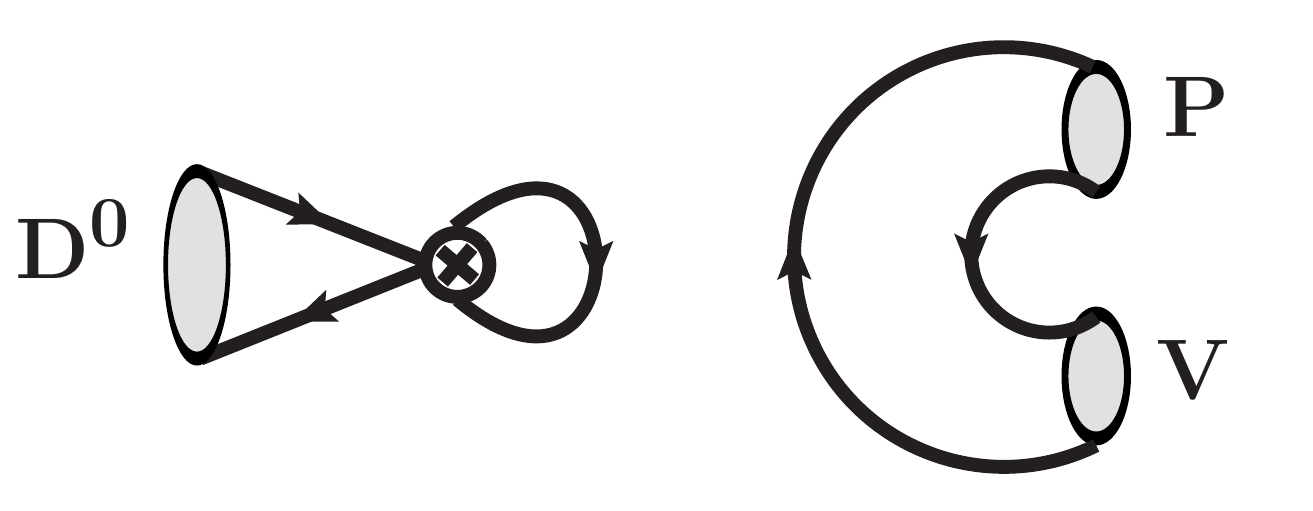}
}
\end{center}
\caption{
$SU(3)_F$-limit topological amplitudes  
 $E_P$ (a),  $E_V$ (b),  $PA_{Pq}$ (c), and $PA_{Vq}$ (d) 
  entering $D^0 \to  \ov{K}{}^0 K^{*0}$ and $D^0\to K^0 \ov{K}{}^{*0}$. 
``$V$'' and ``$P$'' stand for ``vector'' and ``pseudoscalar'', respectively, and   
   label the two different positions of $\ov{K}{}^0$ and $K^{*0}$ 
   in the diagrams.
   The $q$ in $PA_{Pq}$ and $PA_{Vq}$ labels the quark running in the loop at the weak vertex.
   We define $PA_P\equiv PA_{Ps} + PA_{Pd} - 2 PA_{Pb}$ and analogous for $PA_V$. 
   Note that the contributions from $PA_P$ and $PA_V$ cannot be distinguished from each other.
   We use therefore the notation $PA_{PV}\equiv PA_P + PA_V$.
\label{fig:feynman-diags-limit} }
\end{figure*}

Now 
\begin{align}
  \imag \frac{\lambda_b}{\lambda_{sd}} = - 6\cdot 10^{-4} \label{eq:size}
\end{align}
defines the typical size of $|a_{CP}^{\mathrm{dir}}(d)|$.  In
Ref.~\cite{Nierste:2015zra} we have found that
$|a_{CP}^{\mathrm{dir}}(D^0\rightarrow K_SK_S)|$ can be 
as large as $1.1 \cdot 10^{-2}$ and proposed $D^0\rightarrow K_SK_S$
as a discovery channel for charm CP violation. 
Experiments start to probe this region~\cite{Bonvicini:2000qm,Aaij:2015fua, Abdesselam:2016gqq}.
The reason for this enhancement compared to the expectation in \eq{eq:size} is two-fold:
\begin{itemize}
\item[(i)] $|\mathcal{A}_{sd}(D^0\rightarrow K_SK_S)|$ 
     is suppressed, because it vanishes in
     the SU(3)$_F$ symmetry limit, see also Refs.~\cite{Brod:2011re,Atwood:2012ac,Hiller:2012xm}. 
\item[(ii)] $|\mathcal{A}_{b}(D^0\rightarrow K_SK_S)| $ 
     is enhanced, because it involves the large topological 
     amplitude $E$. Contrary to $P$, this amplitude 
      involves no loop (see \fig{fig:feynman-diags-limit}) 
     and a global fit to measured 
     branching ratios supports a large value of $|E|$
     \cite{Muller:2015lua}, comparable to $|T|$. This feature 
     is easily understood, because the color suppression of
     $E$ is offset by a large Wilson coefficient $2C_2\sim 2.4$~\cite{Buras:1985xv}.
\end{itemize}

\begin{figure*}[t]
\begin{center}
\subfigure[\label{}]{
        \includegraphics[width=0.23\textwidth]{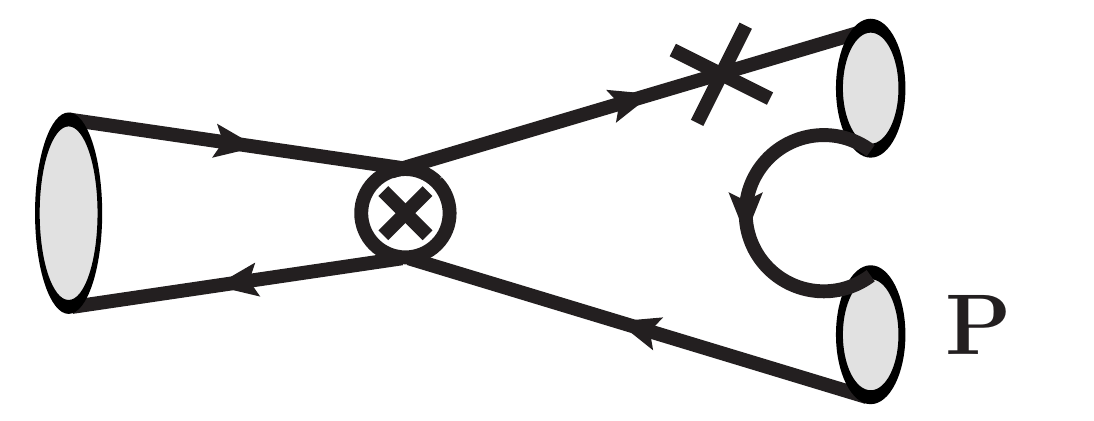}
}
\hfill %
\subfigure[\label{}]{
        \includegraphics[width=0.23\textwidth]{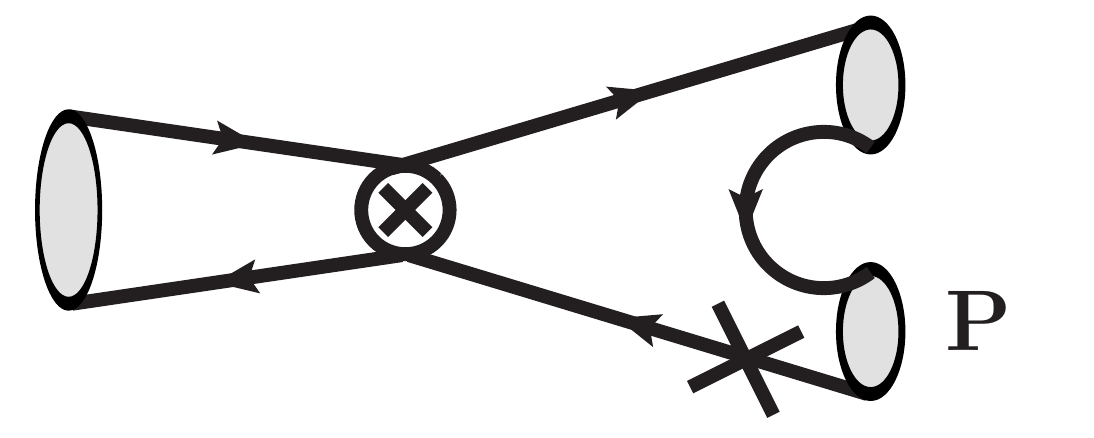}
}
\hfill %
\subfigure[\label{}]{
        \includegraphics[width=0.23\textwidth]{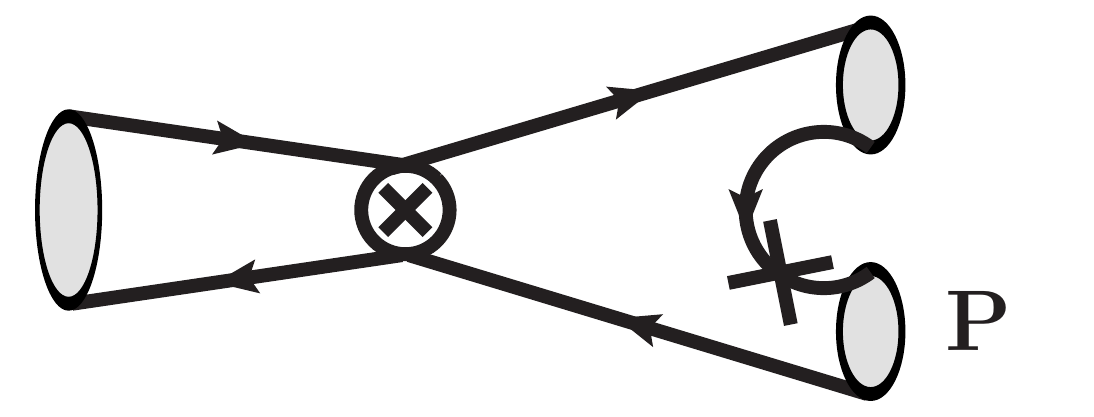}
}
\hfill %
\subfigure[\label{}]{
        \includegraphics[width=0.23\textwidth]{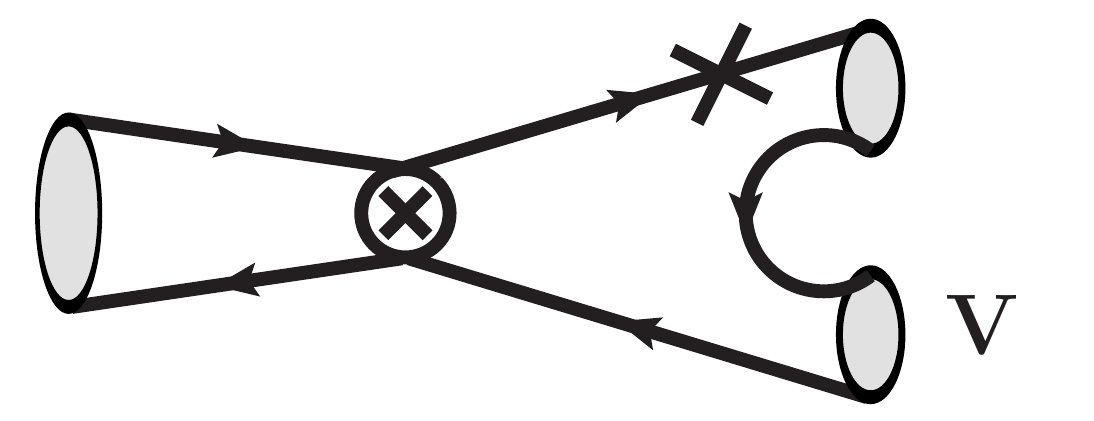}
}
\hfill %
\subfigure[\label{}]{
        \includegraphics[width=0.23\textwidth]{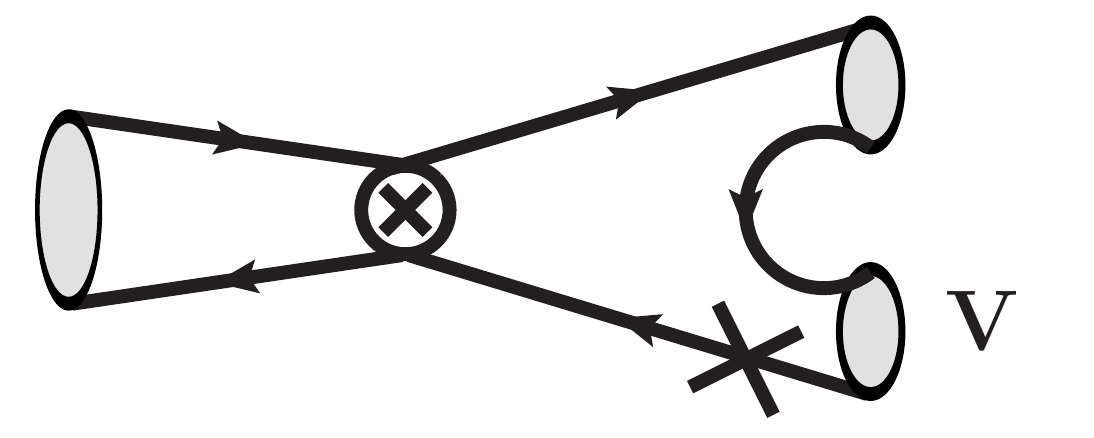}
}
\hfill %
\subfigure[\label{}]{
        \includegraphics[width=0.23\textwidth]{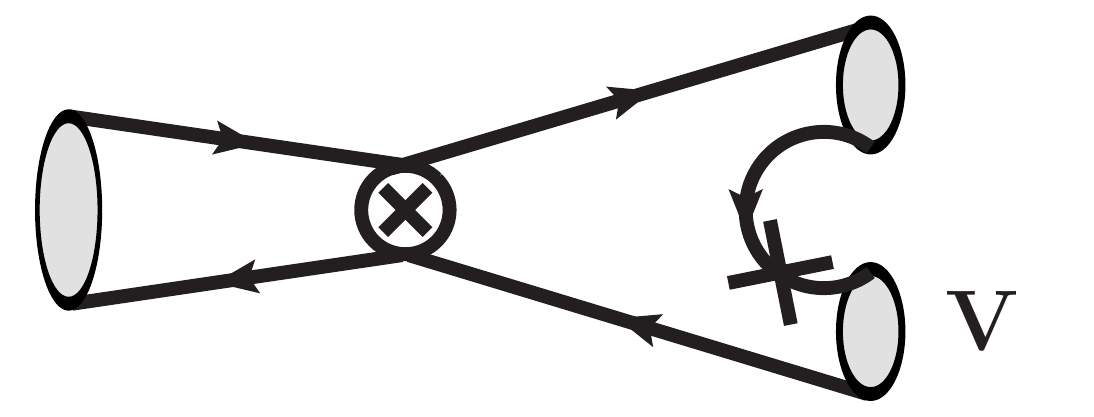}
}
\hfill %
\subfigure[\label{}]{
        \includegraphics[width=0.23\textwidth]{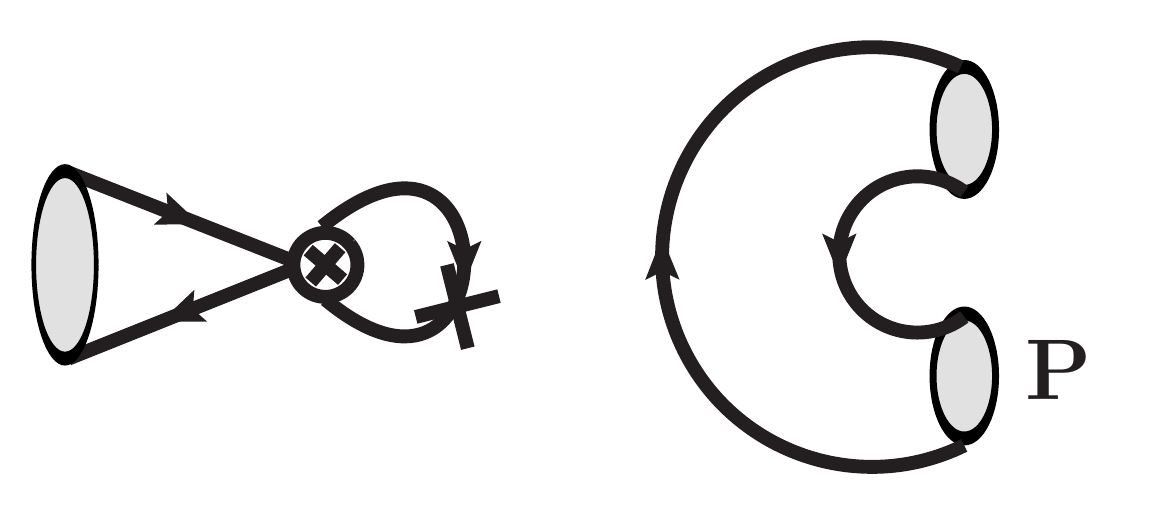}
}
\hfill %
\subfigure[\label{}]{
        \includegraphics[width=0.23\textwidth]{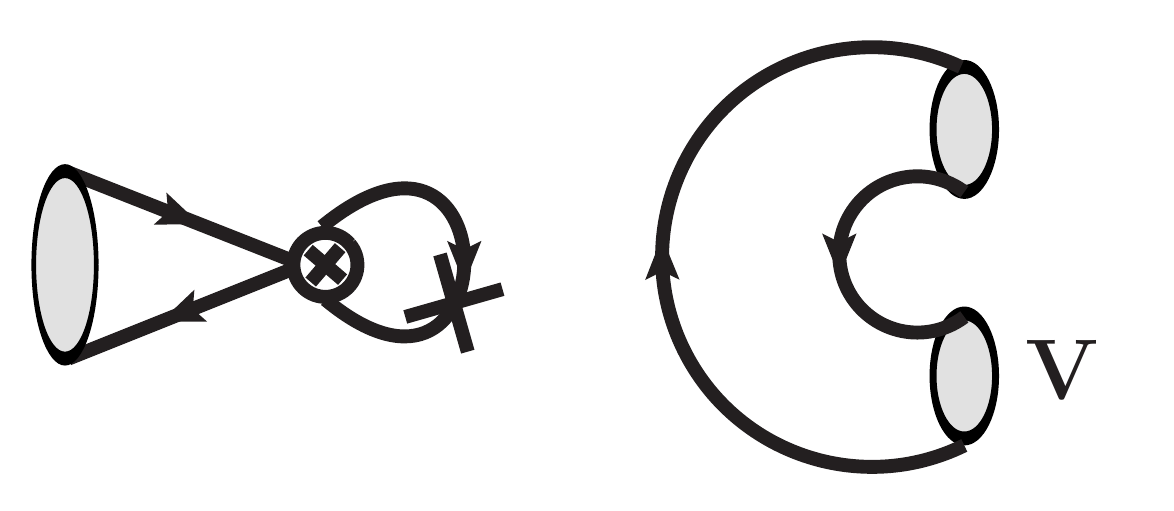}
}
\end{center}
\caption{
$SU(3)_F$-breaking topological amplitudes 
 $E_{P1}$ (a),  $E_{P2}$ (b),  $E_{P3}$ (c), 
 $E_{V1}$ (d) ,  $E_{V2}$ (e),  $E_{V3}$ (f),
 $PA^{\mathrm{break}}_{P}\equiv PA_{Ps} - PA_{Pd}$ (g) and $PA^{\mathrm{break}}_{V} \equiv PA_{Vs} - PA_{Vd}$ (h)
contributing to $D^0 \to  \ov{K}{}^0 K^{*0}$ and $D^0\to K^0 \ov{K}{}^{*0}$.
Note that the contributions from  $PA^{\mathrm{break}}_P$ and $PA^{\mathrm{break}}_V$ cannot be distinguished from each other.
We use therefore the notation $PA^{\mathrm{break}}_{PV}\equiv PA^{\mathrm{break}}_{P} + PA^{\mathrm{break}}_{V}$.
\label{fig:feynman-diags-break}
}
\end{figure*}
 
In this paper we extend the analysis of Ref.~\cite{Nierste:2015zra}
to the decays $D^0 \to  \ov{K}{}^0 K^{*0}$ and $D^0 \to  K^0
\ov{K}{}^{*0}$. The $K^{*0}=K^{*0}(892)$ is understood to be observed 
as $K^{*0}\to K^+\pi^-$, \ie\ in a flavor-specific decay distinguishing 
$K^{*0}$ from  $\ov K{}^{*0}$ decaying as $\ov K{}^{*0}\to K^-\pi^+$.
For the corresponding amplitudes we write 
\begin{align}
\mathcal{A}(\ov{K}{}^{*0}) &\equiv \mathcal{A}(D^0\rightarrow \ov{K}{}^{*0} K^0)  \label{eq:amp-notation-1}\\ 
\mathcal{A}(K^{*0}) 	       &\equiv \mathcal{A}(D^0\rightarrow K^{*0} \ov{K}{}^{0}). \label{eq:amp-notation-2}
\end{align}
At present, these modes are compatible with CP conservation~\cite{Aaij:2015lsa}, however with large errors.
The modes $D^0 \to  \ov{K}{}^0 K^{*0}, K^0 \ov{K}{}^{*0}$ share the
properties (i) and (ii) with $D^0\to K_S K_S$, except that the suppression 
of $|\mathcal{A}_{sd}|$ cannot be inferred from symmetry arguments. 
Instead, the smallness of $|\mathcal{A}_{sd}|$ is only found
empirically, from the branching ratios that we extract from the literature~\cite{Aaij:2015lsa, Olive:2016xmw} as 
\begin{align}
 \mathcal{B}^{\mathrm{exp}} (D^0 \to  K^{*0} K_S) 	 &=  (1.1 \pm 0.2) \cdot 10^{-4}\,,  \label{eq:brr1} \\
 \mathcal{B}^{\mathrm{exp}} (D^0 \to  \ov{K}{}^{*0} K_S) &=  (0.9 \pm 0.2) \cdot 10^{-4}\,.  \label{eq:brr2}
\end{align}
Note that to the given precision, Eqs.~(\ref{eq:brr1}), (\ref{eq:brr2}) do not depend on the choice of the GLASS or LASS scheme in Ref.~\cite{Aaij:2015lsa}.
GLASS and LASS are two models for the $K\pi$ S-wave contributions, see Ref.~\cite{Aaij:2015lsa} for details.
The topological amplitudes contributing to these decays are shown in
\fig{fig:feynman-diags-limit}. Eqs.~(\ref{eq:brr1}) and (\ref{eq:brr2}) entail $E_V\sim E_P$ for the two 
exchange amplitudes, while global fits to the branching ratios
of $D$ decays into a pseudoscalar and a vector meson show that 
$|E_V|$ and $|E_P|$ are individually large, 
with ratios of exchange over tree diagrams between 
0.2 and 0.5~\cite{Bhattacharya:2008ke, Bhattacharya:2012fz, Cheng:2016ejf}.  
For a dedicated discussion of the rates and phases of $D\rightarrow KK^*$ as well 
as comparisons to BaBar~\cite{Aubert:2002yc} and Belle~\cite{Insler:2012pm} Dalitz plot data see  
Refs.~\cite{Bhattacharya:2011gf, Bhattacharya:2012fz}. 
In addition to (i) and (ii) there are more features  
making $D^0 \to  \ov{K}{}^0 K^{*0}, K^0 \ov{K}{}^{*0}$ interesting 
for the hunt for charm CP violation:
\begin{itemize}
\item[(iii)] The prompt decay $K^{*0}\rightarrow K^+\pi^-$ 
    produces charged tracks pointing directly to the $D^0$ decay vertex
    and the problem with the sizable $K_S$ lifetime in $D^0\to K_SK_S$ 
    is alleviated. Unlike the phase-space suppressed decay $D^0 \to  \ov{K}{}^{*0} K^{*0}$ the proposed
    modes require no angular analysis. 
\item[(iv)] Direct CP asymmetries vanish if
  $\mathcal{A}_b/\mathcal{A}_{sd}$ is real, \ie\ if the relative
  strong phase of the interfering amplitudes equals zero or $\pi$.
  Thus to discover CP violation one must be lucky with the uncalculable 
  strong phases. However, in the analysis of the $(K^+,\pi^-,K_S)$
  Dalitz plot one can relax the requirement
  $M(K^+,\pi^-)=M_{K^{0*}}=892\mev$ and scan over invariant masses 
  $M(K^+,\pi^-)$ in the vicinity of the $K^{0*}$ mass, exploiting that
  strong phases strongly vary in the vicinity of resonances.  
\item[(v)] The CP asymmetry does not vanish in the untagged $D^0$
  decay,~\ie\ the decay rates of  
  $\DorDbar \to \ov{K}{}^0 K^{*0}$ 
  and   $\DorDbar \to K^0 \ov{K}{}^{*0}$, 
  differ from each other. 
  Thus no flavor tagging is needed. 
\end{itemize}
We define the untagged rates  
$\Gamma(\DorDbar \to f) \equiv \Gamma(D^0\to f) + \Gamma(\ov{D}{}^0 \to f)$
and obtain the direct CP asymmetry of the untagged $D^0$ decay as
\begin{align}
& a_{CP}^{\mathrm{dir, untag}}(K^{*0}) \equiv \frac{
\Gamma(\DorDbar \to \ov{K}{}^0 K^{*0}) - \Gamma(\DorDbar \to K^0 \ov{K}{}^{*0})
}{
\Gamma(\DorDbar \to \ov{K}{}^0 K^{*0}) + \Gamma(\DorDbar \to K^0 \ov{K}{}^{*0})
} \\
&=
	\mathrm{Im}\frac{\lambda_b}{\lambda_{sd}} 
	\frac{
		\mathrm{Im}\left( 
			\mathcal{A}_{sd}^*(K^{*0}) \mathcal{A}_b(K^{*0}) -
 			\mathcal{A}_{sd}^*(\ov{K}{}^{*0}) \mathcal{A}_b(\ov{K}{}^{*0})
		\right)
		}{
			\vert\mathcal{A}_{sd}(K^{*0})\vert^2 + \vert\mathcal{A}_{sd}(\ov{K}{}^{*0})\vert^2
		} \label{eq:aCPuntagged} \\
&= - a_{CP}^{\mathrm{dir, untag}}(\ov{K}{}^{*0})\,. 
\end{align}
Below, we give the topological decompositions of $\mathcal{A}(K^{*0})$ and $\mathcal{A}(\ov{K}{}^{*0})$, respectively. Subsequently, we insert these into the expressions for the CP asymmetries. 
We analyze the phenomenological implications of the results and conclude.

\section{Topological Decomposition\label{sec:topo}}

Similar in this respect to $\mathcal{A}(D^0\rightarrow K_SK_S)$, the topological decompositions of $\mathcal{A}(\ov{K}{}^{*0})$ and 
$\mathcal{A}(K^{*0})$, see Eqs.~(\ref{eq:amp-notation-1}) and (\ref{eq:amp-notation-2}), depend on exchange and penguin annihilation topologies only:
\begin{align}
\mathcal{A}_{sd}(K^{*0}) &= E_P - E_V \nn\\&\quad + E_{P3} - E_{V1} - E_{V2} - PA^{\mathrm{break}}_{PV}\,, \label{eq:topo-first}\\
\mathcal{A}_b(K^{*0}) &= -E_P-E_V \nn\\&\quad -  E_{P3} - E_{V1} - E_{V2} - PA_{PV} \\
		      &= \mathcal{A}_{sd}( K^{*0} ) - 2 E_P - 2 E_{P3}\nn\\&\quad - PA_{PV} + PA^{\mathrm{break}}_{PV}\,, \label{eq:AbKstar}\\
\mathcal{A}_{sd}(\ov{K}{}^{*0}) &= -E_P + E_V\nn\\&\quad - E_{P1} - E_{P2} + E_{V3} - PA^{\mathrm{break}}_{PV}\,, \label{eq:topo-second} \\ 
\mathcal{A}_b(\ov{K}{}^{*0}) &= -E_P-E_V \nn\\&\quad - E_{P1} - E_{P2} - E_{V3} - PA_{PV} \\
		      &= \mathcal{A}_{sd}(\ov{K}{}^{*0}) - 2 E_V - 2 E_{V3} \nn\\&\quad - PA_{PV} + PA^{\mathrm{break}}_{PV}\,. \label{eq:AbKbarstar}
\end{align}
Note that we express $\mathcal{A}_b$ by $\mathcal{A}_{sd}$ in order to make the subsequent topological dependences of the CP asymmetry more transparent, analogous to Refs.~\cite{Muller:2015rna, Nierste:2015zra}.
Furthermore, we differentiate exchange and penguin annihilation 
diagrams where the antiquark from the weak vertex goes into the pseudoscalar meson ($E_P$, $PA_P$) or 
into the vector meson ($E_V$, $PA_V$). The exact naming scheme for the topologies is defined in 
Figs.~\ref{fig:feynman-diags-limit} and \ref{fig:feynman-diags-break}.
The SU(3)$_F$ limit of Eqs.~(\ref{eq:topo-first})--(\ref{eq:AbKbarstar}) agrees with Ref.~\cite{Cheng:2012wr}, the 
CKM-leading SU(3)$_F$ limit also with Ref.~\cite{Bhattacharya:2008ke}.
We use the amplitude normalization~\cite{Bhattacharya:2012fz}
\begin{align}
\vert \mathcal{A}(D\rightarrow VP)\vert &= 
\sqrt{
	\frac{ 
		 8 \pi m_D^2 \, \mathcal{B}(D\rightarrow VP)
	}{
		\tau_D (p^*)^3
	}
}\,,
\end{align}
with the $D^0$ lifetime $\tau_D$ and $p^*$ the magnitude of the $K_S$, 
$ \raisebox{7.7pt}{$\scriptscriptstyle(\hspace*{8.5pt})$} \hspace*{-10.7pt}\! \overline{K}{}^{\,*}$ 
3-momentum.
For the kaon states we use the conventions $K_S=\frac{1}{\sqrt{2}}(K^0-\ov{K}{}^0)$ and 
$K_L=\frac{1}{\sqrt{2}}(K^0+\ov{K}{}^0)$.\footnote{We assume that effects of kaon CP violation are eliminated with the formula of Ref.~\cite{Grossman:2011zk}.}
For the amplitudes it follows
\begin{align}
\mathcal{A}(D^0\rightarrow K^{*0} K_{S,L}) 	      &= \mp\frac{1}{\sqrt{2}}\mathcal{A}(D^0\rightarrow K^{*0} \ov{K}{}^0)\,,  \label{eq:KSKL-1}\\
\mathcal{A}(D^0\rightarrow \ov{K}{}^{*0} K_{S,L}) &= \frac{1}{\sqrt{2}}\mathcal{A}(D^0\rightarrow \ov{K}{}^{*0} K^0)\,,  \label{eq:KSKL-2}
\end{align}
so that we have for the direct CP asymmetries with tagged charm flavor 
\begin{align}
a_{CP}^{\mathrm{dir}}(D^0\rightarrow K^{*0} K_{S})            &= a_{CP}^{\mathrm{dir}}(D^0\rightarrow K^{*0} K_{L}) \\
							      &=  a_{CP}^{\mathrm{dir}}(D^0\rightarrow K^{*0} \ov{K}{}^0)  \,, \\
a_{CP}^{\mathrm{dir}}(D^0\rightarrow \ov{K}{}^{*0} K_{S}) &= a_{CP}^{\mathrm{dir}}(D^0\rightarrow \ov{K}{}^{*0} K_{L}) \\
							      &= a_{CP}^{\mathrm{dir}}(D^0\rightarrow \ov{K}{}^{*0} K^0) \,.
\end{align}
We write therefore shortly 
\begin{align}
	a_{CP}^{\mathrm{dir}}(K^{*0})		 &\equiv a_{CP}^{\mathrm{dir}}(D^0\rightarrow K^{*0} K_{S})\,, \\
	a_{CP}^{\mathrm{dir}}(\ov{K}{}^{*0}) &\equiv a_{CP}^{\mathrm{dir}}(D^0\rightarrow \ov{K}{}^{*0} K_{S})\,.
\end{align}
Inserting the topological parametrizations Eqs.~(\ref{eq:topo-first})--(\ref{eq:AbKbarstar}) into Eq.~(\ref{eq:aCPdirFormula}) we arrive at
\begin{align}
a_{CP}^{\mathrm{dir}}(K^{*0}) 		 &= - R(K^{*0}) \sin\delta(K^{*0})\,, \label{eq:CPasym-result-1}\\
a_{CP}^{\mathrm{dir}}(\ov{K}{}^{*0}) &= - R(\ov{K}{}^{*0}) \sin\delta(\ov{K}{}^{*0})\,,
\end{align}
with the magnitudes 
\begin{align}
R(K^{*0}) &\equiv -\mathrm{Im}(\lambda_b) / \vert \mathcal{A}(K^{*0})\vert \times\nn\\
		 &\quad \vert -2 (E_P + E_{P3}) - PA_{PV} + PA^{\mathrm{break}}_{PV} \vert\,, \\
R(\ov{K}{}^{*0}) &\equiv -\mathrm{Im}(\lambda_b) / \vert \mathcal{A}(\ov{K}{}^{*0}) \vert \times\nn\\
		&\quad \vert -2 (E_V + E_{V3}) - PA_{PV} + PA^{\mathrm{break}}_{PV} \vert\,,
\end{align}
and the phases
\begin{align}
\delta( K^{*0}) 	   &= \mathrm{arg}\left(
					\frac{
						-2 ( E_P + E_{P3}) - PA_{PV} + PA^{\mathrm{break}}_{PV}
					}{
						\mathcal{A}_{sd}(K^{*0})
					}\right)\,, \label{eq:phase-1}\\ 
\delta( \ov{K}{}^{*0}) &= \mathrm{arg}\left(
					\frac{
						-2 (E_V + E_{V3}) - PA_{PV} + PA^{\mathrm{break}}_{PV}
					}{
						\mathcal{A}_{sd}(\ov{K}{}^{*0})
					}\right)\,. \label{eq:phase-2}
\end{align}
It is instructive to study the SU(3)$_F$ limit of the above expressions. 
To begin with, in the SU(3)$_F$ limit Eqs.~(\ref{eq:topo-first})-(\ref{eq:AbKbarstar}) imply 
\begin{align}
\mathcal{A}_{sd}(K^{*0}) &= -\mathcal{A}_{sd}(\ov{K}{}^{*0})\,, \label{eq:prediction-br} \\
\mathcal{A}_b(K^{*0})    &= \mathcal{A}_b(\ov{K}{}^{*0})\,. \label{eq:su3limit-Ab}
\end{align}
Eq.~(\ref{eq:prediction-br}) agrees with Refs.~\cite{Bhattacharya:2011gf, Bhattacharya:2012fz}.
Although in Eqs.~(\ref{eq:topo-first}), (\ref{eq:topo-second}) several SU(3)$_F$-breaking topologies are present, which in principle 
could affect Eq.~(\ref{eq:prediction-br}) considerably, the latest LHCb data entail~\cite{Aaij:2015lsa}
\begin{align}
\left|\frac{\mathcal{A}( D^0\rightarrow K_S K^{*0} ) }{ \mathcal{A}(D^0\rightarrow K_S \ov{K}{}^{*0} ) } \right|
	&= \begin{cases}
		1.12\pm 0.05\pm 0.11 &\text{(GLASS)} \\
		1.17\pm 0.04\pm 0.05 &\text{(LASS)}
	   \end{cases}\,, \label{eq:LHCb-result}  
\end{align}
meaning small SU(3)$_F$ breaking. 
In the SU(3)$_F$ limit we have 
\begin{align}
a_{CP}^{\mathrm{dir}}(K^{*0}) 
	&=\frac{\mathrm{Im}(\lambda_b)}{\lambda_{sd}} \mathrm{Im}\left( \frac{-2 E_P-PA_{PV}}{E_P - E_V}\right) \\
	&=-\frac{\mathrm{Im}(\lambda_b)}{\lambda_{sd}} \mathrm{Im}\left( \frac{E_P+E_V+PA_{PV}}{E_P - E_V}\right)\,, \label{eq:aCP-1}
\end{align}
and analogously
\begin{align}
a_{CP}^{\mathrm{dir}}(\ov{K}{}^{*0}) 
	&= \frac{\mathrm{Im}(\lambda_b)}{\lambda_{sd}} \mathrm{Im}\left( \frac{E_P+E_V+PA_{PV}}{E_P-E_V}\right)\,, \label{eq:aCP-2}
\end{align} 
showing that $a_{CP}^{\mathrm{dir}}$ is enhanced for $E_P\sim E_V$. 
In the step to Eq.~(\ref{eq:aCP-1}) we added $(E_P-E_V)/(E_P-E_V)$ to the term in brackets.
Eqs.~(\ref{eq:aCP-1}), (\ref{eq:aCP-2}) imply the sum rule
\begin{align}
a_{CP}^{\mathrm{dir}}(K^{*0}) + a_{CP}^{\mathrm{dir}}(\ov{K}{}^{*0}) &= 0\,, \label{eq:approx-sum-rule}
\end{align}
found in Refs.~\cite{Grossman:2012ry, Grossman:2013lya}, which also complies with the numerical results of Ref.~\cite{Cheng:2012wr}.
Eq.~(\ref{eq:approx-sum-rule}) is a test of SU(3)$_F$ breaking in the CP asymmetries, sensitive to other topological amplitudes 
than Eq.~(\ref{eq:prediction-br}).

For the untagged CP asymmetry we arrive  at 
\begin{align}
a_{CP}^{\mathrm{dir,untag}}(K^{*0}) &= a_{CP}^{\mathrm{dir}}(K^{*0}) \\
= -a_{CP}^{\mathrm{dir,untag}}(\ov{K}{}^{*0}) &= -a_{CP}^{\mathrm{dir}}(\ov{K}{}^{*0})  
\end{align}
in the SU(3)$_F$ limit, \ie\ there is no dilution of the untagged CP asymmetry with respect to the tagged one. 
Barring the possibility of accidentally vanishing strong phases, 
$a_{CP}^{\mathrm{dir}}(K^{*0})$ and $a_{CP}^{\mathrm{dir}}(\ov{K}{}^{*0})$
neither vanish in the SU(3)$_F$ limit nor in the limit of vanishing penguin annihilation.
On the contrary, following the above discussion one can expect that the main contribution to the CP asymmetry stems in 
fact from the SU(3)$_F$-limit exchange diagrams $E_P$, $E_V$.

\section{Phenomenology \label{sec:pheno}}

From the LHCb measurements~Eqs.~(\ref{eq:brr1}) and (\ref{eq:brr2}) we extract the absolute value of the difference of the 
exchange topologies as:
\begin{align}
\vert E_P - E_V\vert &= (1.6 \pm 0.2) \cdot 10^{-6}\,. \label{eq:br-constr} 
\end{align}
We use this bound together with the solution for the absolute values of $E_P$ and $E_V$ in Table~1 of Ref.~\cite{Bhattacharya:2012fz},
\begin{align}
\vert E_P\vert &= (2.94 \pm 0.09)\cdot 10^{-6}\,, \label{eq:ep-constr} \\
\vert E_V\vert &= (2.37 \pm 0.19)\cdot 10^{-6}\,. \label{eq:ev-constr} 
\end{align}
For a rough estimate of $a_{CP}^{\text{dir, untag}}$ near the $K^*$ peak we use Eq.~(\ref{eq:aCP-1}) where we vary
$\vert E_P\vert$ and $\vert E_V\vert$ flat inside the $2\sigma$ ranges of Eqs.~(\ref{eq:ep-constr})--(\ref{eq:ev-constr}), while 
imposing the branching ratio constraint Eq.~(\ref{eq:br-constr}) to be also fulfilled at $2\sigma$.
Furthermore, we use $0 \leq \vert PA_{PV}\vert \leq  0.2 \times (E_P+E_V)/2$ with the central values of $E_P$, $E_V$ in Eqs.~(\ref{eq:ep-constr}, \ref{eq:ev-constr}). 
All relative  strong phases are varied freely. The constraints from 
Eqs.~(\ref{eq:br-constr})--(\ref{eq:ev-constr}) pin the relevant relative strong phase between $E_P$ and $E_V$
down to the interval $[-0.24\pi, +0.24\pi]$.  
The maximum value of $a_{CP}^{\text{dir, untag}}$ near the peak of the $K^*$~resonance is then 
\begin{align}
\vert a_{CP}^{\text{dir, untag}}\vert \lesssim 0.003\,, \label{eq:aCPnum-peak}
\end{align}
with the maximum found for $\mathrm{arg}(E_V/E_P) = 0.14\,\pi$. 
In the experimental hunt for charm CP discovery one can further scan the Dalitz plot around 
the $K^*$~resonance to look for favourable strong phases which maximize $\vert a_{CP}^{\text{dir, untag}}\vert$.

In order to inspect the dependence of this result on the size of penguin annihilation diagrams we also 
look at the case $PA_{PV} = 0$. 
As the dominant piece of the CP asymmetry stems from the exchange topologies, 
we find the result in Eq.~(\ref{eq:aCPnum-peak}) unchanged.

\section{Conclusions \label{sec:conclusion}} 

CP asymmetries in neutral $D\rightarrow KK^*$ decays are driven by exchange topologies and persist in the 
limit of vanishing penguins. 
In the SU(3)$_F$ limit the untagged CP asymmetry is equal to the tagged one,~\ie\ there is no dilution, which enables the 
search for charm CP violation with high statistics in untagged samples.
Therefore $D\rightarrow KK^*$ decays are promising discovery channels for charm CP violation.
Our estimate for the maximum possible CP asymmetry is given in Eq.~(\ref{eq:aCPnum-peak}).

\begin{acknowledgments}
We thank Olli Lupton, Brian Meadows and Guy Wilkinson for advice on the extraction of the branching ratio data Eqs.~(\ref{eq:brr1}) and (\ref{eq:brr2}) from 
Ref.~\cite{Aaij:2015lsa,Olive:2016xmw}. We thank Michael Morello and Fu-Sheng Yu for useful discussions.
The presented work is supported by BMBF under contract no.~05H15VKKB1. 
\end{acknowledgments}

\bibliography{kkstar.bib}

\begin{thebibliography}{22}%
\makeatletter
\providecommand \@ifxundefined [1]{%
 \@ifx{#1\undefined}
}%
\providecommand \@ifnum [1]{%
 \ifnum #1\expandafter \@firstoftwo
 \else \expandafter \@secondoftwo
 \fi
}%
\providecommand \@ifx [1]{%
 \ifx #1\expandafter \@firstoftwo
 \else \expandafter \@secondoftwo
 \fi
}%
\providecommand \natexlab [1]{#1}%
\providecommand \enquote  [1]{``#1''}%
\providecommand \bibnamefont  [1]{#1}%
\providecommand \bibfnamefont [1]{#1}%
\providecommand \citenamefont [1]{#1}%
\providecommand \href@noop [0]{\@secondoftwo}%
\providecommand \href [0]{\begingroup \@sanitize@url \@href}%
\providecommand \@href[1]{\@@startlink{#1}\@@href}%
\providecommand \@@href[1]{\endgroup#1\@@endlink}%
\providecommand \@sanitize@url [0]{\catcode `\\12\catcode `\$12\catcode
  `\&12\catcode `\#12\catcode `\^12\catcode `\_12\catcode `\%12\relax}%
\providecommand \@@startlink[1]{}%
\providecommand \@@endlink[0]{}%
\providecommand \url  [0]{\begingroup\@sanitize@url \@url }%
\providecommand \@url [1]{\endgroup\@href {#1}{\urlprefix }}%
\providecommand \urlprefix  [0]{URL }%
\providecommand \Eprint [0]{\href }%
\providecommand \doibase [0]{http://dx.doi.org/}%
\providecommand \selectlanguage [0]{\@gobble}%
\providecommand \bibinfo  [0]{\@secondoftwo}%
\providecommand \bibfield  [0]{\@secondoftwo}%
\providecommand \translation [1]{[#1]}%
\providecommand \BibitemOpen [0]{}%
\providecommand \bibitemStop [0]{}%
\providecommand \bibitemNoStop [0]{.\EOS\space}%
\providecommand \EOS [0]{\spacefactor3000\relax}%
\providecommand \BibitemShut  [1]{\csname bibitem#1\endcsname}%
\let\auto@bib@innerbib\@empty
\bibitem [{\citenamefont {Nierste}\ and\ \citenamefont
  {Schacht}(2015)}]{Nierste:2015zra}%
  \BibitemOpen
  \bibfield  {author} {\bibinfo {author} {\bibfnamefont {U.}~\bibnamefont
  {Nierste}}\ and\ \bibinfo {author} {\bibfnamefont {S.}~\bibnamefont
  {Schacht}},\ }\href {\doibase 10.1103/PhysRevD.92.054036} {\bibfield
  {journal} {\bibinfo  {journal} {Phys. Rev.}\ }\textbf {\bibinfo {volume}
  {D92}},\ \bibinfo {pages} {054036} (\bibinfo {year} {2015})},\ \Eprint
  {http://arxiv.org/abs/1508.00074} {arXiv:1508.00074 [hep-ph]} \BibitemShut
  {NoStop}%
\bibitem [{\citenamefont {Bonvicini}\ \emph {et~al.}(2001)\citenamefont
  {Bonvicini} \emph {et~al.}}]{Bonvicini:2000qm}%
  \BibitemOpen
  \bibfield  {author} {\bibinfo {author} {\bibfnamefont {G.}~\bibnamefont
  {Bonvicini}} \emph {et~al.} (\bibinfo {collaboration} {CLEO Collaboration}),\
  }\href {\doibase 10.1103/PhysRevD.63.071101} {\bibfield  {journal} {\bibinfo
  {journal} {Phys.Rev.}\ }\textbf {\bibinfo {volume} {D63}},\ \bibinfo {pages}
  {071101} (\bibinfo {year} {2001})},\ \Eprint
  {http://arxiv.org/abs/hep-ex/0012054} {arXiv:hep-ex/0012054 [hep-ex]}
  \BibitemShut {NoStop}%
\bibitem [{\citenamefont {Aaij}\ \emph {et~al.}(2015)\citenamefont {Aaij} \emph
  {et~al.}}]{Aaij:2015fua}%
  \BibitemOpen
  \bibfield  {author} {\bibinfo {author} {\bibfnamefont {R.}~\bibnamefont
  {Aaij}} \emph {et~al.} (\bibinfo {collaboration} {LHCb}),\ }\href {\doibase
  10.1007/JHEP10(2015)055} {\bibfield  {journal} {\bibinfo  {journal} {JHEP}\
  }\textbf {\bibinfo {volume} {10}},\ \bibinfo {pages} {055} (\bibinfo {year}
  {2015})},\ \Eprint {http://arxiv.org/abs/1508.06087} {arXiv:1508.06087
  [hep-ex]} \BibitemShut {NoStop}%
\bibitem [{\citenamefont {Abdesselam}\ \emph {et~al.}(2016)\citenamefont
  {Abdesselam} \emph {et~al.}}]{Abdesselam:2016gqq}%
  \BibitemOpen
  \bibfield  {author} {\bibinfo {author} {\bibfnamefont {A.}~\bibnamefont
  {Abdesselam}} \emph {et~al.},\ }\href@noop {} {\  (\bibinfo {year} {2016})},\
  \Eprint {http://arxiv.org/abs/1609.06393} {arXiv:1609.06393 [hep-ex]}
  \BibitemShut {NoStop}%
\bibitem [{\citenamefont {Brod}\ \emph {et~al.}(2012)\citenamefont {Brod},
  \citenamefont {Kagan},\ and\ \citenamefont {Zupan}}]{Brod:2011re}%
  \BibitemOpen
  \bibfield  {author} {\bibinfo {author} {\bibfnamefont {J.}~\bibnamefont
  {Brod}}, \bibinfo {author} {\bibfnamefont {A.~L.}\ \bibnamefont {Kagan}}, \
  and\ \bibinfo {author} {\bibfnamefont {J.}~\bibnamefont {Zupan}},\ }\href
  {\doibase 10.1103/PhysRevD.86.014023} {\bibfield  {journal} {\bibinfo
  {journal} {Phys.Rev.}\ }\textbf {\bibinfo {volume} {D86}},\ \bibinfo {pages}
  {014023} (\bibinfo {year} {2012})},\ \Eprint {http://arxiv.org/abs/1111.5000}
  {arXiv:1111.5000 [hep-ph]} \BibitemShut {NoStop}%
\bibitem [{\citenamefont {Atwood}\ and\ \citenamefont
  {Soni}(2013)}]{Atwood:2012ac}%
  \BibitemOpen
  \bibfield  {author} {\bibinfo {author} {\bibfnamefont {D.}~\bibnamefont
  {Atwood}}\ and\ \bibinfo {author} {\bibfnamefont {A.}~\bibnamefont {Soni}},\
  }\href {\doibase 10.1093/ptep/ptt065} {\bibfield  {journal} {\bibinfo
  {journal} {PTEP}\ }\textbf {\bibinfo {volume} {2013}},\ \bibinfo {pages}
  {0903B05} (\bibinfo {year} {2013})},\ \Eprint
  {http://arxiv.org/abs/1211.1026} {arXiv:1211.1026 [hep-ph]} \BibitemShut
  {NoStop}%
\bibitem [{\citenamefont {Hiller}\ \emph {et~al.}(2013)\citenamefont {Hiller},
  \citenamefont {Jung},\ and\ \citenamefont {Schacht}}]{Hiller:2012xm}%
  \BibitemOpen
  \bibfield  {author} {\bibinfo {author} {\bibfnamefont {G.}~\bibnamefont
  {Hiller}}, \bibinfo {author} {\bibfnamefont {M.}~\bibnamefont {Jung}}, \ and\
  \bibinfo {author} {\bibfnamefont {S.}~\bibnamefont {Schacht}},\ }\href
  {\doibase 10.1103/PhysRevD.87.014024} {\bibfield  {journal} {\bibinfo
  {journal} {Phys.Rev.}\ }\textbf {\bibinfo {volume} {D87}},\ \bibinfo {pages}
  {014024} (\bibinfo {year} {2013})},\ \Eprint {http://arxiv.org/abs/1211.3734}
  {arXiv:1211.3734 [hep-ph]} \BibitemShut {NoStop}%
\bibitem [{\citenamefont {{M\"uller}}\ \emph
  {et~al.}(2015{\natexlab{a}})\citenamefont {{M\"uller}}, \citenamefont
  {Nierste},\ and\ \citenamefont {Schacht}}]{Muller:2015lua}%
  \BibitemOpen
  \bibfield  {author} {\bibinfo {author} {\bibfnamefont {S.}~\bibnamefont
  {{M\"uller}}}, \bibinfo {author} {\bibfnamefont {U.}~\bibnamefont {Nierste}},
  \ and\ \bibinfo {author} {\bibfnamefont {S.}~\bibnamefont {Schacht}},\ }\href
  {\doibase 10.1103/PhysRevD.92.014004} {\bibfield  {journal} {\bibinfo
  {journal} {Phys. Rev.}\ }\textbf {\bibinfo {volume} {D92}},\ \bibinfo {pages}
  {014004} (\bibinfo {year} {2015}{\natexlab{a}})},\ \Eprint
  {http://arxiv.org/abs/1503.06759} {arXiv:1503.06759 [hep-ph]} \BibitemShut
  {NoStop}%
\bibitem [{\citenamefont {Buras}\ \emph {et~al.}(1986)\citenamefont {Buras},
  \citenamefont {Gerard},\ and\ \citenamefont {Ruckl}}]{Buras:1985xv}%
  \BibitemOpen
  \bibfield  {author} {\bibinfo {author} {\bibfnamefont {A.}~\bibnamefont
  {Buras}}, \bibinfo {author} {\bibfnamefont {J.}~\bibnamefont {Gerard}}, \
  and\ \bibinfo {author} {\bibfnamefont {R.}~\bibnamefont {Ruckl}},\ }\href
  {\doibase 10.1016/0550-3213(86)90200-2} {\bibfield  {journal} {\bibinfo
  {journal} {Nucl.Phys.}\ }\textbf {\bibinfo {volume} {B268}},\ \bibinfo
  {pages} {16} (\bibinfo {year} {1986})}\BibitemShut {NoStop}%
\bibitem [{\citenamefont {Aaij}\ \emph {et~al.}(2016)\citenamefont {Aaij} \emph
  {et~al.}}]{Aaij:2015lsa}%
  \BibitemOpen
  \bibfield  {author} {\bibinfo {author} {\bibfnamefont {R.}~\bibnamefont
  {Aaij}} \emph {et~al.} (\bibinfo {collaboration} {LHCb}),\ }\href {\doibase
  10.1103/PhysRevD.93.052018} {\bibfield  {journal} {\bibinfo  {journal} {Phys.
  Rev.}\ }\textbf {\bibinfo {volume} {D93}},\ \bibinfo {pages} {052018}
  (\bibinfo {year} {2016})},\ \Eprint {http://arxiv.org/abs/1509.06628}
  {arXiv:1509.06628 [hep-ex]} \BibitemShut {NoStop}%
\bibitem [{\citenamefont {Patrignani}\ \emph {et~al.}(2016)\citenamefont
  {Patrignani} \emph {et~al.}}]{Olive:2016xmw}%
  \BibitemOpen
  \bibfield  {author} {\bibinfo {author} {\bibfnamefont {C.}~\bibnamefont
  {Patrignani}} \emph {et~al.} (\bibinfo {collaboration} {Particle Data
  Group}),\ }\href {\doibase 10.1088/1674-1137/40/10/100001} {\bibfield
  {journal} {\bibinfo  {journal} {Chin. Phys.}\ }\textbf {\bibinfo {volume}
  {C40}},\ \bibinfo {pages} {100001} (\bibinfo {year} {2016})}\BibitemShut
  {NoStop}%
\bibitem [{\citenamefont {Bhattacharya}\ and\ \citenamefont
  {Rosner}(2009)}]{Bhattacharya:2008ke}%
  \BibitemOpen
  \bibfield  {author} {\bibinfo {author} {\bibfnamefont {B.}~\bibnamefont
  {Bhattacharya}}\ and\ \bibinfo {author} {\bibfnamefont {J.~L.}\ \bibnamefont
  {Rosner}},\ }\href {\doibase 10.1103/PhysRevD.79.034016,
  10.1103/PhysRevD.81.099903} {\bibfield  {journal} {\bibinfo  {journal}
  {Phys.Rev.}\ }\textbf {\bibinfo {volume} {D79}},\ \bibinfo {pages} {034016}
  (\bibinfo {year} {2009})},\ \Eprint {http://arxiv.org/abs/0812.3167}
  {arXiv:0812.3167 [hep-ph]} \BibitemShut {NoStop}%
\bibitem [{\citenamefont {Bhattacharya}\ and\ \citenamefont
  {Rosner}(2012)}]{Bhattacharya:2012fz}%
  \BibitemOpen
  \bibfield  {author} {\bibinfo {author} {\bibfnamefont {B.}~\bibnamefont
  {Bhattacharya}}\ and\ \bibinfo {author} {\bibfnamefont {J.~L.}\ \bibnamefont
  {Rosner}},\ }\href {\doibase 10.1016/j.physletb.2012.07.009} {\bibfield
  {journal} {\bibinfo  {journal} {Phys. Lett.}\ }\textbf {\bibinfo {volume}
  {B714}},\ \bibinfo {pages} {276} (\bibinfo {year} {2012})},\ \Eprint
  {http://arxiv.org/abs/1203.6014} {arXiv:1203.6014 [hep-ph]} \BibitemShut
  {NoStop}%
\bibitem [{\citenamefont {Cheng}\ \emph {et~al.}(2016)\citenamefont {Cheng},
  \citenamefont {Chiang},\ and\ \citenamefont {Kuo}}]{Cheng:2016ejf}%
  \BibitemOpen
  \bibfield  {author} {\bibinfo {author} {\bibfnamefont {H.-Y.}\ \bibnamefont
  {Cheng}}, \bibinfo {author} {\bibfnamefont {C.-W.}\ \bibnamefont {Chiang}}, \
  and\ \bibinfo {author} {\bibfnamefont {A.-L.}\ \bibnamefont {Kuo}},\ }\href
  {\doibase 10.1103/PhysRevD.93.114010} {\bibfield  {journal} {\bibinfo
  {journal} {Phys. Rev.}\ }\textbf {\bibinfo {volume} {D93}},\ \bibinfo {pages}
  {114010} (\bibinfo {year} {2016})},\ \Eprint
  {http://arxiv.org/abs/1604.03761} {arXiv:1604.03761 [hep-ph]} \BibitemShut
  {NoStop}%
\bibitem [{\citenamefont {Aubert}\ \emph {et~al.}(2002)\citenamefont {Aubert}
  \emph {et~al.}}]{Aubert:2002yc}%
  \BibitemOpen
  \bibfield  {author} {\bibinfo {author} {\bibfnamefont {B.}~\bibnamefont
  {Aubert}} \emph {et~al.} (\bibinfo {collaboration} {BaBar}),\ }in\ \href
  {https://oraweb.slac.stanford.edu/pls/slacquery/BABAR_DOCUMENTS.Search?P_SLAC_PUB=SLAC-PUB-9320}
  {\emph {\bibinfo {booktitle} {{High energy physics. Proceedings, 31st
  International Conference, ICHEP 2002, Amsterdam, Netherlands, July 25-31,
  2002}}}}\ (\bibinfo {year} {2002})\ \Eprint
  {http://arxiv.org/abs/hep-ex/0207089} {arXiv:hep-ex/0207089 [hep-ex]}
  \BibitemShut {NoStop}%
\bibitem [{\citenamefont {Insler}\ \emph {et~al.}(2012)\citenamefont {Insler}
  \emph {et~al.}}]{Insler:2012pm}%
  \BibitemOpen
  \bibfield  {author} {\bibinfo {author} {\bibfnamefont {J.}~\bibnamefont
  {Insler}} \emph {et~al.} (\bibinfo {collaboration} {CLEO}),\ }\href {\doibase
  10.1103/PhysRevD.85.092016} {\bibfield  {journal} {\bibinfo  {journal} {Phys.
  Rev.}\ }\textbf {\bibinfo {volume} {D85}},\ \bibinfo {pages} {092016}
  (\bibinfo {year} {2012})},\ \Eprint {http://arxiv.org/abs/1203.3804}
  {arXiv:1203.3804 [hep-ex]} \BibitemShut {NoStop}%
\bibitem [{\citenamefont {Bhattacharya}\ and\ \citenamefont
  {Rosner}(2011)}]{Bhattacharya:2011gf}%
  \BibitemOpen
  \bibfield  {author} {\bibinfo {author} {\bibfnamefont {B.}~\bibnamefont
  {Bhattacharya}}\ and\ \bibinfo {author} {\bibfnamefont {J.~L.}\ \bibnamefont
  {Rosner}},\ }\href@noop {} {\  (\bibinfo {year} {2011})},\ \Eprint
  {http://arxiv.org/abs/1104.4962} {arXiv:1104.4962 [hep-ph]} \BibitemShut
  {NoStop}%
\bibitem [{\citenamefont {{M\"uller}}\ \emph
  {et~al.}(2015{\natexlab{b}})\citenamefont {{M\"uller}}, \citenamefont
  {Nierste},\ and\ \citenamefont {Schacht}}]{Muller:2015rna}%
  \BibitemOpen
  \bibfield  {author} {\bibinfo {author} {\bibfnamefont {S.}~\bibnamefont
  {{M\"uller}}}, \bibinfo {author} {\bibfnamefont {U.}~\bibnamefont {Nierste}},
  \ and\ \bibinfo {author} {\bibfnamefont {S.}~\bibnamefont {Schacht}},\ }\href
  {\doibase 10.1103/PhysRevLett.115.251802} {\bibfield  {journal} {\bibinfo
  {journal} {Phys. Rev. Lett.}\ }\textbf {\bibinfo {volume} {115}},\ \bibinfo
  {pages} {251802} (\bibinfo {year} {2015}{\natexlab{b}})},\ \Eprint
  {http://arxiv.org/abs/1506.04121} {arXiv:1506.04121 [hep-ph]} \BibitemShut
  {NoStop}%
\bibitem [{\citenamefont {Cheng}\ and\ \citenamefont
  {Chiang}(2012)}]{Cheng:2012wr}%
  \BibitemOpen
  \bibfield  {author} {\bibinfo {author} {\bibfnamefont {H.-Y.}\ \bibnamefont
  {Cheng}}\ and\ \bibinfo {author} {\bibfnamefont {C.-W.}\ \bibnamefont
  {Chiang}},\ }\href {\doibase 10.1103/PhysRevD.85.079903,
  10.1103/PhysRevD.85.034036} {\bibfield  {journal} {\bibinfo  {journal}
  {Phys.Rev.}\ }\textbf {\bibinfo {volume} {D85}},\ \bibinfo {pages} {034036}
  (\bibinfo {year} {2012})},\ \Eprint {http://arxiv.org/abs/1201.0785}
  {arXiv:1201.0785 [hep-ph]} \BibitemShut {NoStop}%
\bibitem [{\citenamefont {Grossman}\ and\ \citenamefont
  {Nir}(2012)}]{Grossman:2011zk}%
  \BibitemOpen
  \bibfield  {author} {\bibinfo {author} {\bibfnamefont {Y.}~\bibnamefont
  {Grossman}}\ and\ \bibinfo {author} {\bibfnamefont {Y.}~\bibnamefont {Nir}},\
  }\href {\doibase 10.1007/JHEP04(2012)002} {\bibfield  {journal} {\bibinfo
  {journal} {JHEP}\ }\textbf {\bibinfo {volume} {1204}},\ \bibinfo {pages}
  {002} (\bibinfo {year} {2012})},\ \Eprint {http://arxiv.org/abs/1110.3790}
  {arXiv:1110.3790 [hep-ph]} \BibitemShut {NoStop}%
\bibitem [{\citenamefont {Grossman}\ and\ \citenamefont
  {Robinson}(2013)}]{Grossman:2012ry}%
  \BibitemOpen
  \bibfield  {author} {\bibinfo {author} {\bibfnamefont {Y.}~\bibnamefont
  {Grossman}}\ and\ \bibinfo {author} {\bibfnamefont {D.~J.}\ \bibnamefont
  {Robinson}},\ }\href {\doibase 10.1007/JHEP04(2013)067} {\bibfield  {journal}
  {\bibinfo  {journal} {JHEP}\ }\textbf {\bibinfo {volume} {1304}},\ \bibinfo
  {pages} {067} (\bibinfo {year} {2013})},\ \Eprint
  {http://arxiv.org/abs/1211.3361} {arXiv:1211.3361 [hep-ph]} \BibitemShut
  {NoStop}%
\bibitem [{\citenamefont {Grossman}\ \emph {et~al.}(2014)\citenamefont
  {Grossman}, \citenamefont {Ligeti},\ and\ \citenamefont
  {Robinson}}]{Grossman:2013lya}%
  \BibitemOpen
  \bibfield  {author} {\bibinfo {author} {\bibfnamefont {Y.}~\bibnamefont
  {Grossman}}, \bibinfo {author} {\bibfnamefont {Z.}~\bibnamefont {Ligeti}}, \
  and\ \bibinfo {author} {\bibfnamefont {D.~J.}\ \bibnamefont {Robinson}},\
  }\href {\doibase 10.1007/JHEP01(2014)066} {\bibfield  {journal} {\bibinfo
  {journal} {JHEP}\ }\textbf {\bibinfo {volume} {1401}},\ \bibinfo {pages}
  {066} (\bibinfo {year} {2014})},\ \Eprint {http://arxiv.org/abs/1308.4143}
  {arXiv:1308.4143 [hep-ph]} \BibitemShut {NoStop}%
\end{thebibliography}%

\end{document}